\documentclass[12pt]{article}

\usepackage{amsmath}
\usepackage{amsfonts}
\usepackage[nosort]{cite}

\numberwithin{equation}{section}

\def\beq#1\eeq{\begin{equation}#1\end{equation}}
\def\bes#1\ees{\begin{equation}\begin{split}#1\end{split}\end{equation}}
\def\bea#1\eea{\begin{align}#1\end{align}}

\newcommand{\Z}{\mathbb{Z}}

\newcommand{\bra}[1]{\langle #1|}
\newcommand{\ket}[1]{|#1\rangle}

\newcommand{\dbra}[1]{\langle\!\langle #1|}
\newcommand{\dket}[1]{|#1 \rangle\!\rangle}

\newcommand{\auto}{\mathcal{O}}
\newcommand{\fusion}[4][\mathcal{N}]{{#1}_{#2 #3}{}^{#4}}
\newcommand{\components}[3]{{#1}_{#2}{}^{#3}}
\newcommand{\E}{\mathcal{E}}
\newcommand{\V}{\mathcal{V}}

\DeclareMathOperator{\Spec}{Spec}
\begin{document}
\baselineskip=6mm

%%%%%%%%%%%%%%%%%%%%%%%%%%%%%%%%%%%%%%%%%%%%%%%%%%%%%%%%
%%   Titlepage
%%%%%%%%%%%%%%%%%%%%%%%%%%%%%%%%%%%%%%%%%%%%%%%%%%%%%%%%
\begin{titlepage}
\nopagebreak
\vskip 5mm
\begin{flushright}
hep-th/0111230 \\
TU-640
\end{flushright}

\vskip 10mm
\begin{center}
{\Large \textbf{Boundary states in coset conformal field theories}}
\vskip 15mm
Hiroshi \textsc{Ishikawa}
\vskip 5mm
\textsl{%
Department of Physics, Tohoku University \\
Sendai 980-8578, JAPAN\\
}
\texttt{%
\footnotesize
ishikawa@tuhep.phys.tohoku.ac.jp
}
\end{center}
\vskip 15mm

%%%%%%%%%%%%%%%%%%%% Abstract %%%%%%%%%%%%%%%%%%%%%%%%%%
\begin{quote}
We construct various boundary states in the coset conformal field theory
$G/H$.
The $G/H$ theory admits the twisted boundary condition if the $G$ theory
has an outer automorphism of the horizontal subalgebra that induces
an automorphism of the $H$ theory.
By introducing the notion of the brane identification and the brane
selection rule, we show that the twisted boundary states of the $G/H$ theory
can be constructed from those of the $G$ and the $H$ theories.
We apply our construction to the $su(n)$ diagonal cosets
and the $su(2)/u(1)$ parafermion theory to obtain the twisted
boundary states of these theories.
\end{quote}

\vfill
\end{titlepage}

%%%%%%%%%%%%%%%%%%%%%%%%%%%%%%%%%%%%%%%%%%%%%%%%%%%%%%%%
%%   Intorduction
%%%%%%%%%%%%%%%%%%%%%%%%%%%%%%%%%%%%%%%%%%%%%%%%%%%%%%%%
\section{Introduction}
The recent developments in the construction of the boundary states
in rational conformal field theories have revealed the rich structure
of conformal field theories with boundaries \cite{FS, BFS, BPPZ}.
It is now recognized that the rational boundary states are
described by
a non-negative integer matrix
representation (NIM-rep) of the fusion algebra \cite{DZ,BPPZ,Gannon}.
The situation changes, however, if we relax the condition of rationality
on the boundary states. 
Many rational CFT's are equipped with an extended symmetry such as the current
algebra, and they are no more rational with respect to the Virasoro algebra.
If we require only the conformal invariance on the boundary states,
instead of the full chiral algebra, the classification problem
gets much complicated, for which we have no generic answer. 

In the context of string theory, boundary states give rise to
D-branes. 
In order to have a consistent theory, D-branes have to keep
(super) conformal invariance on the worldsheet.
The conservation of the extended current algebra is an additional requirement,
which is not necessary in general.
The study of the conformal boundary states is therefore
inevitable for the full understanding of the spectrum of D-branes.

An interesting approach to the construction of the conformal boundary
states in the WZW models has been proposed in \cite{MMS}
\footnote{%
For the $c=1$ models, it is possible to take another
approach \cite{GRW,GR,Janik}. 
}. 
The strategy of \cite{MMS} is to decompose the $G$ WZW model 
into the $H$ part and the coset $G/H$, where $H$ is a subgroup of
the group $G$
\beq
  G \sim G/H \times H .
\eeq
From this decomposition, we have several boundary conditions of the $G$
theory.
Adopting the usual boundary condition for both of the $H$ and the $G/H$
parts yields the ordinary boundary condition of the $G$ theory.
On the other hand,
we can twist the boundary condition of the $H$ part by
an automorphism that leaves the Virasoro algebra invariant. 
Twisting the $G/H$ part in the same way does not affect the boundary
condition of the $G$ theory.
Taking the ordinary condition in the coset theory, however, gives
the novel boundary condition of the $G$ theory.
This condition breaks the $G$ current algebra while its conformal invariance
is manifest. 
In \cite{MMS}, 
it has been shown that 
the conformal boundary state, not rational with respect to $G$, does exist 
for the case of $G = SU(2)$.
The boundary states in the coset theories are therefore useful building blocks
in the construction of the conformal boundary states in the WZW models.
Although the coset theory with boundaries has been studied from the sigma model
point of view \cite{MMS,G,ES,FScho2}, 
the algebraic study such as \cite{FS,BFS,BPPZ} is necessary 
to explore the stringy regime of the theory.

In this paper, we give the general method to obtain
the boundary states in the $G/H$ coset conformal theory. 
In particular, we show that a NIM-rep of
the $G/H$ theory can be constructed from a pair of NIM-reps for the
$G$ and the $H$ theories.
In doing this, we introduce the notion of the brane identification
and the brane selection rule,
which are considered to be the boundary version of the field identification
and the selection rule in the coset theory.
We apply our method to
the twisted boundary states of 
the $su(n)_1 \oplus su(n)_1/su(n)_2$ diagonal
coset and the $su(2)_k/u(1)_k$ parafermion theory,
and obtain the result consistent with that in \cite{MMS}.

The organization of the paper is as follows.
In the next section, we review some results about the boundary
states in rational conformal field theories, especially the WZW models. 
In Section 3, we give arguments for the existence of an automorphism of
the boundary states,
which is the dual of the automorphism of the current algebra. 
In Section 4, we give the rule to yield NIM-reps in
the coset theory.
We show that a pair of NIM-reps in the $G$ and the $H$ theories
yield a NIM-rep in the $G/H$ theory after an appropriate
identification of the states generated by the automorphism of the boundary
states.
In Section 5, we apply our method to several examples.

%%%%%%%%%%%%%%%%%%%%%%%%%%%%%%%%%%%%%%%%%%%%%%%%%%%%%%%%
%%   WZW models
%%%%%%%%%%%%%%%%%%%%%%%%%%%%%%%%%%%%%%%%%%%%%%%%%%%%%%%%
\section{Boundary states in the WZW models}

%%%%%%%%%%%%%%%%%%%%%%%%%%%%%%%%%%%%%%%%%%%%%%%%%%%%%%%%
%\subsection{Cardy states}
In this section, we review some basic results about the boundary states 
in the WZW models following to \cite{FS,BFS,BPPZ}.

The most simple boundary condition for the current algebra is
\beq
\label{symmetric_bc}
  J^a_n + \tilde{J}^a_{-n} = 0 ,
\eeq
where $J^a$ and $\tilde{J}^a$ represent the holomorphic and the anti-holomorphic
parts of the algebra, respectively.
The Ishibashi states $\{\dket{\lambda}\, |\, \lambda \in \Spec(G)\}$ 
are the building blocks of the boundary states \cite{Ishibashi}.
Here, we denote by $\Spec(G)$ the set of the integrable representations
of the algebra $g$ at level $k$, namely, 
$\Spec(G) = P_+^k(g)$. 
We normalize $\dket{\lambda}$ as follows
\beq
  \dbra{\lambda} \tilde{q}^{H_c} \dket{\lambda} 
  = \frac{1}{S_{0\lambda}} \chi_\lambda(-1/\tau) 
  = \sum_{\mu \in \Spec(G)}
    \frac{S_{\lambda \mu}}{S_{0 \lambda}} \chi_\mu(\tau)
  = \chi_0(\tau) + \cdots ,
\eeq
where $\tilde{q} = e^{-2\pi i /\tau}$ and
$H_c = \frac{1}{2}(L_0 + \tilde{L}_0 - \frac{c}{12})$ 
is the closed string Hamiltonian.
`$0$' stands for the vacuum representation.
This normalization corresponds to the following scalar product 
in the space of the boundary states \cite{BPPZ}
\beq
  \bra{\alpha} \ket{\beta} =
  \lim_{q \rightarrow 0} q^{\frac{c}{24}} 
  \bra{\alpha} \tilde{q}^{H_c} \ket{\beta} .
\eeq
Here $c$ is the central charge of the theory and $q = e^{2\pi i \tau}$.

The boundary condition \eqref{symmetric_bc} relates a representation
$\lambda$ with $\bar{\lambda}$. 
Hence, $(\lambda)_L \otimes (\bar{\lambda})_R$ must exist 
in the closed string spectrum in order to have the Ishibashi state
$\dket{\lambda}$. 
In the case of the charge-conjugation modular invariant
$Z = \sum_{\lambda \in \Spec(G)} \chi_\lambda \bar{\chi}_{\bar{\lambda}}$,
we obtain all the Ishibashi states $\dket{\lambda}, \lambda \in \Spec(G)$.
However, in the other cases, the set of the allowed Ishibashi states is
in general different from $\Spec(G)$. We denote this set by $\E$
\beq
\label{E}
  \E = \{\lambda \,|\,
        (\lambda)_L \otimes (\bar{\lambda})_R \in
\text{closed string spectrum} \}.
\eeq
For the diagonal modular invariant
$Z = \sum_{\lambda \in \Spec(G)} \chi_\lambda \bar{\chi}_{\lambda}$,
only the self-conjugate representations are allowed and
$\E = \{\lambda \in \Spec(G)\,|\, \bar{\lambda} = \lambda\}$. 
The multiplicity of a representation $\lambda$ in $\E$ can be greater than 1,
as is seen in the $D_{\text{even}}$ invariant of $su(2)$.

%%%%%%%%%
A generic boundary state $\ket{\alpha}$ satisfying the boundary condition
\eqref{symmetric_bc} is a linear combination of
the Ishibashi states
\beq
  \ket{\alpha} = \sum_{\lambda \in \E}
  \components{\psi}{\alpha}{\lambda} \dket{\lambda}.
\eeq
We denote by $\V$ the set labelling the boundary states
\beq
  \V = \{\alpha \,|\, \text{label of the boundary states}\} .
\eeq
The annulus amplitude between two boundary states takes the form
\beq
  Z_{\alpha \beta} = \bra{\beta} \tilde{q}^{H_c} \ket{\alpha} 
  = \sum_{\lambda \in \E, \mu \in \Spec(G)}
  \components{\psi}{\alpha}{\lambda} 
  \frac{S_{\mu \lambda}}{S_{0 \lambda}}
  \components{\bar{\psi}}{\beta}{\lambda}\, \chi_\mu(\tau) 
  = \sum_{\mu \in \Spec(G)} \fusion[n]{\mu}{\alpha}{\beta} \chi_\mu .
\eeq
Here we denote the multiplicity of the representation $\mu$ in $Z_{\alpha \beta}$
by $\fusion[n]{\mu}{\alpha}{\beta}$
\begin{subequations}
\label{NIM}
\beq
  \fusion[n]{\mu}{\alpha}{\beta} = 
  \sum_{\lambda \in \E}
  \components{\psi}{\alpha}{\lambda} 
  \frac{S_{\mu \lambda}}{S_{0 \lambda}}
  \components{\bar{\psi}}{\beta}{\lambda} =
  \sum_{\lambda \in \E}
  \components{\psi}{\alpha}{\lambda} 
  \gamma^{(\mu)}_\lambda
  \components{\bar{\psi}}{\beta}{\lambda} ,
\eeq
In the matrix form, this can be written as
\beq
  n_\mu = \psi  \gamma^{(\mu)} \psi^\dagger ,
\eeq
\end{subequations}
where $\components{(\psi)}{\alpha}{\lambda} = \components{\psi}{\alpha}{\lambda}$
and $\components{(n_\mu)}{\alpha}{\beta} = \fusion[n]{\mu}{\alpha}{\beta}$.
Here we denote by $\gamma^{(\mu)}_\lambda$ the generalized quantum dimension
\beq
\label{quantum_dim}
  \gamma^{(\lambda)} = \text{diag}(\gamma^{(\lambda)}_{\;\rho}) =
  \text{diag}\left(\frac{S_{\lambda \rho}}{S_{0 \rho}}\right)_{\rho \in P_+^k} .
\eeq 
Clearly, $\fusion[n]{\mu}{\alpha}{\beta}$ takes non-negative integer values
for the consistent boundary conditions \cite{Cardy}.
Moreover, $\fusion[n]{0}{\alpha}{\beta} = \delta_{\alpha\beta}$ since
the vacuum is unique, and 
$n_\mu^T$ is related with $n_{\bar{\mu}}$ via
\beq
  n_\mu^T = n_\mu^\dagger = (\psi  \gamma^{(\mu)} \psi^\dagger)^\dagger
  = \psi  (\gamma^{(\mu)})^\dagger \psi^\dagger
  = \psi  \gamma^{(\bar{\mu})} \psi^\dagger
  = n_{\bar{\mu}} .
\eeq
We call the set of the consistency conditions the Cardy condition
\beq
\label{Cardy}
  \fusion[n]{\mu}{\alpha}{\beta} \in \Z_{\ge 0} , \quad
  \fusion[n]{0}{\alpha}{\beta} = \delta_{\alpha \beta} \;\;
  (n_0 = 1) , \quad
  \fusion[n]{\mu}{\beta}{\alpha}
    = \fusion[n]{\bar{\mu}}{\alpha}{\beta} \;\;
  (n_\mu^T = n_{\bar{\mu}}) ,
\eeq
and the boundary states satisfying the Cardy condition
the Cardy states. 
It should be noted that the Cardy condition is only a necessary condition
for consistency. 
There are many non-physical NIM-reps that do not correspond
to any modular invariant \cite{Gannon} (\textit{e.g.},
the tadpole NIM-rep of $su(2)$ \cite{DZ,BPPZ}).

So far, the number of the independent Cardy states $\lvert \V \rvert$
is not specified.
From now on, we \textit{assume} that the number of the Cardy states is equal to
the number of the Ishibashi states \cite{PSS}
\beq
\label{completeness}
  \lvert \V \rvert = \lvert \E \rvert \quad 
  \text{(assumption of completeness)} .
\eeq
In other words, the boundary state coefficient $\psi$ is
a square matrix. 
From the Cardy condition \eqref{Cardy},
$n_0 = \psi \psi^\dagger = 1$. For a square $\psi$, this means that
$\psi$ is unitary. 
The situation is quite analogous to the Verlinde formula \cite{Verlinde}
\begin{subequations}
\label{Verlinde}
\beq
  \fusion{\lambda}{\mu}{\nu}
  = \sum_{\rho \in \Spec(G)} 
    \frac{S_{\lambda \rho} S_{\mu \rho} \bar{S}_{\nu \rho}}{S_{0 \rho}}
  = \sum_{\rho \in \Spec(G)} 
    S_{\mu \rho} \gamma^{(\lambda)}_\rho \bar{S}_{\nu \rho},
\eeq
where $\fusion{\lambda}{\mu}{\nu}$ is the fusion coefficient
$(\lambda) \times (\mu) = \sum_{\nu} \fusion{\lambda}{\mu}{\nu} (\nu)$.
In the matrix form, this can be written as
\beq
  N_\lambda = S \gamma^{(\lambda)} S^\dagger ,
\eeq
\end{subequations}
where $\components{(N_\lambda)}{\mu}{\nu} = \fusion{\lambda}{\mu}{\nu}$.
From the associativity of the fusion algebra, one can show that
$N_\lambda$ satisfies the fusion algebra
\beq
  N_\lambda N_\mu = \sum_{\nu \in \Spec(G)} \fusion{\lambda}{\mu}{\nu} N_\nu ,
\eeq
which implies by the Verlinde formula that
\beq
  \gamma^{(\lambda)} \gamma^{(\mu)} = 
  \sum_{\nu \in \Spec(G)} \fusion{\lambda}{\mu}{\nu} \gamma^{(\nu)} .
\eeq
The generalized quantum dimension $\{\gamma^{(\lambda)}_\mu\}$ is therefore
a one-dimensional representation of the fusion algebra.
If we use $\psi$ instead of $S$ in the Verlinde formula,
we obtain $n_\mu$. 
Hence, $n_\mu$, as well as $N_\mu$, satisfies the fusion algebra
\beq
  n_\lambda n_\mu = \sum_{\nu \in \Spec(G)} \fusion{\lambda}{\mu}{\nu} n_\nu .
\eeq
The Cardy condition \eqref{Cardy} together with
the assumption of completeness \eqref{completeness} implies that
$\{n_\lambda\}$ forms a non-negative integer matrix representation
(NIM-rep) of the fusion algebra.

For each set of the mutually consistent boundary states,
we have a NIM-rep of the fusion algebra.
However, the converse is in general not true. 
There are many `unphysical' NIM-reps that do not correspond
to any modular invariant \cite{Gannon}.
The typical example is the tadpole NIM-rep $T_n$ of $su(2)_{2n-1}$
\cite{DZ,BPPZ}, which can be constructed by orbifolding
the regular NIM-rep $A_{2n}$.
The exponent $\E(T_n)$ consists of only the even representations
of $su(2)$ at level $2n-1$.
Hence, there is no modular invariant compatible with $\E(T_n)$
since the level is odd.
Although the spectrum of the diagonal modular invariant at level $2n-1$
contains $\E(T_n)$ as a subset, the overlap of the $T_n$ boundary states
with the ordinary $A_{2n}$ yields the $su(2)$ character with
irrational coefficients, which implies that the $T_n$ state is unphysical.
This example shows that the Cardy condition is not a sufficient
condition for consistency.

We give some examples of the Cardy states below.

%%%%%%%%%%%%%%%%%%%%%%%%%%%%%%%%%%%%%%%%%%%%%%%%%%%%%%%%
\subsection{Untwisted states}
Since the non-negative integer matrix
$\components{(N_\lambda)}{\mu}{\nu} = \fusion{\lambda}{\mu}{\nu}$
satisfies the fusion algebra, $\{N_\lambda\}$ is a NIM-rep of the fusion
algebra (regular representation). 
From the Verlinde formula \eqref{Verlinde},
the corresponding diagonalization matrix $\psi$ is the modular transformation
matrix $S$.
The spectrum $\E$ of the Ishibashi states coincides with $\Spec(G)$.
Hence the resulting Cardy states are those for the charge conjugation
modular invariant.
Since the $S$-matrix maps $\Spec(G)$ to $\Spec(G)$ itself,
the label $\V$ of the Cardy states also coincides with $\Spec(G)$.
The Cardy states take the form \cite{Cardy}
\beq
  \ket{\lambda} = \sum_{\mu \in \Spec(G)} S_{\lambda \mu} \dket{\mu} .
\eeq

%%%%%%%%%%%%%%%%%%%%%%%%%%%%%%%%%%%%%%%%%%%%%%%%%%%%%%%%
\subsection{Twisted states}

The simple Lie algebra $g$ has an outer automorphism $\omega$
for $g = A_l, D_l, E_6$
(see \tablename~\ref{tab:omega}).
%%%%%%%%%%%%%%%%%%%%%%%%
\begin{table}[b]
\caption{The diagram automorphism $\omega$ of
the simple Lie algebra $g$. $r$ is the order of $\omega$ and
$\{\lambda_1, \lambda_2, \cdots \}$
is the Dynkin label of the weight $\lambda$.}
\label{tab:omega}
\begin{center}
\begin{tabular}{|c|c|c|c|c|} \hline
$g$    &  $\omega(\lambda)$         & $r$  \\ \hline
$A_{2l}$  &  $(\lambda_{2l}, \lambda_{2l-1}, \cdots, \lambda_1)$ 
          & $2$   \\
$A_{2l-1}$  &  $(\lambda_{2l-1}, \lambda_{2l-2}, \cdots, \lambda_1)$ 
          & $2$   \\ %\hline
$D_{l+1}$  &  $(\lambda_1, \cdots, \lambda_{l-1}, \lambda_{l+1}, \lambda_l)$ 
          & $2$   \\ %\hline
$E_6$  &  $(\lambda_5, \lambda_4, \lambda_3, \lambda_2, \lambda_1, \lambda_6)$ 
          & $2$   \\ %\hline
$D_4$  &  $(\lambda_4,\lambda_2,\lambda_1,\lambda_3)$ 
          & $3$   \\ \hline
\end{tabular}
\end{center}
\end{table}
%%%%%%%%%%%%%%%%%%%%%%%%
Here, $r$ is the order of $\omega$ and we denote the representation of $g$
by its Dynkin label, namely,
$\lambda = \lambda_1 \Lambda_1 + \cdots + \lambda_l \Lambda_l$.

We can use this outer automorphism $\omega$ of the horizontal
subalgebra $g$ to twist the boundary condition of the current algebra
$g^{(1)}$
\beq
\label{breaking_bc}
  J^a_n + \omega(\tilde{J^a}_{-n}) = 0 .
\eeq
Since $\lambda \neq \omega(\lambda)$ for a generic representation $\lambda$,
the spectrum $\E$ of the Ishibashi states is restricted to
\beq
  \E = P_+^{k, \omega}(g^{(1)}) =
  \{\lambda \in P_+^{k}(g^{(1)}) \,|\, \omega(\lambda) = \lambda \} . 
\eeq
(for simplicity, we consider only the charge-conjugation modular invariant).
The Cardy states are labelled by the integrable representation of the twisted
affine Lie algebra $g^{(r)}$
associated with $g$ and $\omega$ \cite{BFS}
\beq
  \V = P_+^k(g^{(r)}) .
\eeq
This can be understood as follows. 

Let $\{\dket{\lambda; \omega}\,|\, \omega(\lambda) = \lambda\}$ 
be the Ishibashi states satisfying the boundary
condition \eqref{breaking_bc}.
The Cardy state
$\ket{\alpha; \omega}$ can be expressed in terms of $\dket{\lambda; \omega}$
\beq
  \ket{\alpha; \omega} = \sum_{\lambda \in \E}
  \components{\psi}{\alpha}{\lambda} \dket{\lambda; \omega} .
\eeq 
Consider the annulus amplitude between $\ket{\alpha; \omega}$ and
the untwisted Cardy state
$\ket{0} = \sum_{\lambda \in \Spec(G)} S_{0\lambda} \dket{\lambda}$
\beq
\label{twisted_amplitude}
  Z_{0,(\alpha; \omega)} = \bra{0} \tilde{q}^{H_c} \ket{\alpha; \omega}
  = \sum_{\lambda \in \E} S_{0\lambda} \components{\psi}{\alpha}{\lambda}
    \dbra{\lambda} \tilde{q}^{H_c} \dket{\lambda; \omega} . 
\eeq
In the open string channel, the boundary condition of the current $J^a$
is twisted at the one end of the annulus
that corresponds to $\ket{\alpha; \omega}$.
Hence, the current algebra in the open string channel is twisted
to yield the twisted affine Lie algebra $g^{(r)}$, and the annulus
amplitude can be expressed in terms of the character of $g^{(r)}$.
The modular transformation of the character of $g^{(r)}$ is
those for another twisted algebra $\tilde{g}^{(r)}$ \cite{Kac}
(see \tablename~\ref{tab:modular}).
%%%%%%%%%%%%%%%%%%%%%%%%
\begin{table}[b]
\caption{The modular transformation of the twisted affine Lie algebras}
\label{tab:modular}
\begin{center}
\begin{tabular}{|c|ccccc|} \hline
$g^{(r)}$            &  $A_{2l}^{(2)}$  &  $A_{2l-1}^{(2)}$
&  $D_{l+1}^{(2)}$   &  $E_{6}^{(2)}$   & $D_4^{(3)}$ \\ \hline
$\tilde{g}^{(r)}$    &  $A_{2l}^{(2)}$  &  $D_{l+1}^{(2)}$
&  $A_{2l-1}^{(2)}$  &  $E_{6}^{(2)}$   & $D_{4}^{(3)}$   \\ \hline
\end{tabular}
\end{center}
\end{table}
%%%%%%%%%%%%%%%%%%%%%%%%
The overlap $\dbra{\lambda} \tilde{q}^{H_c} \dket{\lambda; \omega}$
of two Ishibashi states is therefore nothing but the character
of $\tilde{g}^{(r)}$ \cite{FSS}. In our normalization, we obtain
\footnote{
For $A_{2l}^{(2)}$, the arguments of the characters should be
slightly modified.
}
\beq
\label{lambda_tilde}
  \dbra{\lambda} \tilde{q}^{H_c} \dket{\lambda; \omega}
  = \frac{1}{S_{0\lambda}} \chi^{\tilde{g}^{(r)}}_{\tilde{\lambda}}(-1/\tau)
  = \frac{1}{S_{0\lambda}} \sum_{\mu \in P_+^k(g^{(r)})}
    \tilde{S}_{\tilde{\lambda} \mu} 
        \chi^{g^{(r)}}_{\mu}(\tau/r) ,
\eeq
where $\tilde{S}$ is the modular transformation matrix
between $g^{(r)}$ and $\tilde{g}^{(r)}$. 
$\tilde{\lambda} \in P_+^k(\tilde{g}^{(r)})$ is determined from
$\lambda \in P_+^{k, \omega}(g^{(1)})$ by comparing the modular anomaly.
We display the concrete form of $\tilde{\lambda}$ in
\tablename~\ref{tab:twin}.
%%%%%%%%%%%%%
\begin{table}
\begin{center}
\begin{tabular}{|cc|cc|} \hline
$g^{(1)}$            &  $\lambda \in P_+^{k, \omega}(g^{(1)})$  &
$\tilde{g}^{(r)}$    &  $\tilde{\lambda} \in P_+^{k}(\tilde{g}^{(r)})$ \\ \hline
$A_{2l}^{(1)}$       &
$(\lambda_1,\cdots,\lambda_l,\lambda_l,\cdots,\lambda_1)$  &
$A_{2l}^{(2)}$       &
$(\lambda_1,\cdots,\lambda_l)$ \\ 
$A_{2l-1}^{(1)}$     &
$(\lambda_1,\cdots,\lambda_l,\cdots,\lambda_1)$  &
$D_{l+1}^{(2)}$       &
$(\lambda_1,\cdots,\lambda_l)$ \\ 
$D_{l+1}^{(1)}$     &
$(\lambda_1,\cdots,\lambda_l,\lambda_l)$  &
$A_{2l-1}^{(2)}$       &
$(\lambda_1,\cdots,\lambda_l)$ \\
$E_{6}^{(1)}$     &
$(\lambda_1,\lambda_2,\lambda_3,\lambda_2,\lambda_1,\lambda_6)$  &
$E_{6}^{(2)}$       &
$(\lambda_6,\lambda_3,\lambda_2,\lambda_1)$ \\ \hline
$D_{4}^{(1)}$     &
$(\lambda_1,\lambda_2,\lambda_1,\lambda_1)$  &
$D_{4}^{(3)}$       &
$(\lambda_2,\lambda_1)$ \\ \hline
\end{tabular}
\end{center}
\caption{The integrable representation $\tilde{\lambda}$ of
the twisted affine Lie algebra $\tilde{g}^{(r)}$
that corresponds to the untwisted affine Lie algebra $g^{(1)}$ and
the diagram automorphism $\omega$ of the horizontal subalgebra
(see eq.\eqref{lambda_tilde}).}
\label{tab:twin}
\end{table}
%%%%%%%%%%%%%

Using this fact,
the annulus amplitude \eqref{twisted_amplitude} 
can be written in the form
\beq
  Z_{0,(\alpha; \omega)} 
  = \sum_{\tilde{\lambda} \in P_+^k(\tilde{g}^{(r)})} 
    \sum_{\mu \in P_+^k(g^{(r)})} \components{\psi}{\alpha}{\lambda}
    \tilde{S}_{\tilde{\lambda} \mu} 
        \chi^{g^{(r)}}_{\mu} .
\eeq
For the consistent boundary states, the coefficient of the character
$\chi^{g^{(r)}}_{\mu}$ should be non-negative integer.
Clearly, this condition is satisfied by setting
$\components{\psi}{\alpha}{\lambda} = \tilde{S}_{\alpha \tilde{\lambda}}$,
where $\alpha \in P_+^k(g^{(r)})$.
Hence, we set
\beq
\label{twisted_states}
  \ket{\alpha; \omega} = \sum_{\lambda \in P_+^{k, \omega}(g^{(1)})}
  \tilde{S}_{\alpha \tilde{\lambda}} \dket{\lambda; \omega} , \quad
  \alpha \in P_+^k(g^{(r)}) .
\eeq
The consistency with the symmetric states $\ket{\beta}$ 
other than $\ket{0}$ readily follows
since $\ket{\beta}$ can be obtained from $\ket{0}$
by the fusion in the open string channel \cite{Cardy}.
In order to see the mutual consistency of the twisted Cardy states, 
we consider the annulus amplitude between two twisted states
\beq
  Z_{(\alpha; \omega)(\beta; \omega)} = 
  \sum_{\lambda \in P_+^{k, \omega}(g^{(1)})}
  \sum_{\mu \in P_+^k(g^{(1)})}
  \tilde{S}_{\alpha \tilde{\lambda}} \frac{S_{\lambda \mu}}{S_{\lambda 0}}
  (\tilde{S}^\dagger)_{\tilde{\lambda} \beta}\, \chi_\mu 
  = \sum_{\mu} N^{\omega}_{\mu \alpha}{}^{\beta} \chi_{\mu} . 
\eeq
For the consistency of the states, the coefficients 
$N^{\omega}_{\mu \alpha}{}^{\beta}$ should be non-negative integer
(a NIM-rep of the fusion algebra).
One can see that this number also
appears in the annulus amplitude between $\ket{\alpha}$ and
$\ket{\beta; \omega}$
\beq
  Z_{\alpha (\beta; \omega)} = 
  \sum_{\lambda \in P_+^{k, \omega}(g^{(1)})}
  \sum_{\mu \in P_+^k(g^{(2)})}
  S_{\alpha \lambda} \frac{\tilde{S}_{\tilde{\lambda} \mu}}{S_{\lambda 0}}
  (\tilde{S}^\dagger)_{\tilde{\lambda} \beta}\, \chi_\mu 
  = \sum_{\lambda, \mu}
  N^{\omega}_{\alpha \mu}{}^{\beta} \chi_\mu .
\eeq
Hence, the mutual consistency of the twisted states follows from
the consistency between the symmetric and the twisted states.

%%%%%%%%%%%%%%%%%%%%%%%%%%%%%%%%%%%%%%%%%%%%%%%%%%%%%%%%
\subsection{$u(1)_k$}

The $u(1)_k$ chiral algebra with $k \in \Z$ has $2k$ primary fields.
We label them by $m \in \Z/2k\Z$
\beq
  \Spec(u(1)_k) = \Z/2k\Z = \{m = 0, 1, \cdots, 2k-1\} .
\eeq
The modular transformation matrix reads
\beq
\label{S_U1}
  S_{mm'} = \frac{1}{\sqrt{2k}} e^{-\frac{\pi i}{k} m m'} ,
\eeq
and the fusion algebra has the form
$(m) \times (m') = (m+m')$. 

Let us consider the twisted boundary states in this theory.
The outer automorphism of $u(1)$ is the charge conjugation 
$\omega_c : m \rightarrow -m$. 
The representations self-conjugate under $\omega_c$ are $m =0$ and $k$.
Hence, we obtain
\beq
  \E = \{0, k\} .
\eeq
Correspondingly, we have two Cardy states denoted by $\alpha = \pm$
\beq
  \V = \{+, -\} .
\eeq
The boundary state coefficient $\psi$ reads
\beq
  \psi = \frac{1}{\sqrt{2}} \begin{pmatrix} 1 & 1 \\ 1 & -1 \end{pmatrix} .
\eeq

%%%%%%%%%%%%%%%%%%%%%%%%%%%%%%%%%%%%%%%%%%%%%%%%%%%%%%%%
%%   Automorphisms
%%%%%%%%%%%%%%%%%%%%%%%%%%%%%%%%%%%%%%%%%%%%%%%%%%%%%%%%
\section{Automorphisms of boundary states}
Let $\text{Aut}(g)$ be the group of the outer automorphism of
a current algebra $g = g^{(1)}$.
$\text{Aut}(g)$ contains a normal abelian subgroup $\auto(g)$
which is isomorphic to the center $Z(G)$ of the group $G$
\beq
  \auto(g) \simeq Z(G) \, , \quad
  \auto(g) \ni A \mapsto b(A) = e^{-2\pi i A(\Lambda_0)} \in Z(G)  .
\eeq
Here $\Lambda_0$ is the 0-th fundamental weight of $g$ 
and $b$ is a group isomorphism
\beq
  b(A A') = b(A) b(A') \, , \quad A, A' \in \auto(g) .
\eeq
$b(A)$ is a multiple of the identity within an irreducible representation
of $g$. 
We denote the eigenvalue of $b(A)$ in the representation $\lambda$ by
$b_\lambda(A)$
\beq
  b(A) \ket{\lambda} = b_\lambda(A) \ket{\lambda} ,\quad
  b_\lambda(A) = e^{-2\pi i (A(\Lambda_0), \lambda)}, \quad
  \lambda \in P_+^k(g) .
\eeq

The modular transformation matrix
$S$ intertwines $\auto(g)$ with $Z(G)$ \cite{Bernard}
\begin{subequations}
\label{Ab}
\beq
  S_{A\lambda, \mu} = S_{\lambda \mu} b_\mu(A) ,
  \quad A \in \auto(g) ,
\eeq
which can be written in the form
\beq
  A S = S\, b(A),  \quad
  A_{\lambda \mu} = \delta_{A\lambda, \mu}, \quad
  b(A) = \text{diag}(b_\lambda(A)) .
\eeq
\end{subequations}
Setting $\lambda=0$ in this equation, we obtain
\beq
\label{b_gamma}
  b_\mu(A) = \frac{S_{A0, \mu}}{S_{0 \mu}} 
  = \gamma^{(A0)}_\mu .
\eeq
Therefore $b_\mu(A)$ is nothing but the generalized quantum
dimension.
The outer automorphism $A \in \auto(g)$ acts on the fusion algebra as
\begin{subequations}
\beq
  \fusion{A\lambda}{\,\mu}{\nu} = \fusion{\lambda}{\,A\mu}{\nu}
  = \fusion{\lambda}{\mu}{A^{-1}\nu} ,
\eeq
which follows from eq.\eqref{Ab} and the Verlinde formula \eqref{Verlinde}.
In the matrix form, this can be written as
\beq
  N_{A\lambda} = A N_\lambda = N_\lambda A .
\eeq
\end{subequations}
Setting $\lambda = 0$ again, we obtain
\beq
\label{AN}
  A = A N_0 = N_{A 0} .
\eeq
Hence, 
the action of the outer automorphism $A$ is equivalent to
the fusion with $A0$ (simple currents \cite{SY}).
This is the `$S$-dual' of eq.\eqref{b_gamma}. 

The outer automorphism $A \in \auto(g)$ naturally acts on the label of
the Cardy states $\V$,
since $A = N_{A0}$.
We use the same symbol $A$ for its realization on $\V$
\beq
  A = n_{A0} = \psi \gamma^{(A0)} \psi^\dagger 
    = \psi b(A) \psi^\dagger .
\eeq
In the component form,
\beq
  \components{A}{\alpha}{\beta} = \sum_{\lambda \in \E}
  \components{\psi}{\alpha}{\lambda}
  b_{\lambda}(A) \components{\bar{\psi}}{\beta}{\lambda} .
\eeq
We can rewrite this as
\begin{subequations}
\label{Apsi}
\bea
  A \psi &= \psi b(A) ,  \\
  \components{\psi}{A\alpha}{\lambda}
  &\equiv \sum_{\beta \in \E}
   \components{A}{\alpha}{\beta}\components{\psi}{\beta}{\lambda}
   = \components{\psi}{\alpha}{\lambda} b_\lambda(A) ,
\eea
\end{subequations}
where we define $A\alpha \in \V$ by
\beq
  \components{A}{\alpha}{\beta} = \delta_{A\alpha, \beta} .
\eeq
There are elements of $\auto(g)$
that leave $\V$ invariant. They form a subgroup of $\auto(g)$, which we
call the stabilizer of $\V$ and denote by $S(\V)$
\bes
  S(\V) &= 
  \{A \in \auto(g)\,|\, A\alpha = \alpha  \text{ for any } \alpha \in \V \} \\
        &= \{A \in \auto(g)\,|\,
             b_\lambda(A) = 1 \text{ for any } \lambda \in \E\}.
\ees
The action on $\V$ is caused by the quotient group
$\auto(\V) \equiv \auto(g)/S(\V)$,
which we call the automorphism group of $\V$
\beq
  \auto(\V) = \auto(g)/S(\V) .
\eeq

From \eqref{Ab} and $b^T = b$, we obtain
\beq
  b(A) = S A^T S^\dagger .
\eeq
We consider the counterpart of this in $\V$, namely, 
\beq
  \tilde{b}(A) = \psi A^T \psi^\dagger , \quad
  A \in \auto(\E) \subset \auto(g) ,
\eeq
where $\auto(\E)$ is the group of outer automorphisms of $\E$ defined as
\beq
  \auto(\E) = \{A \in \auto(g)\,|\, A(\E) = \E \} .
\eeq
This restriction for $A$ is necessary 
because the indices of $\psi$ runs only $\E$.
It is not clear for the author whether $\tilde{b}(A)$ is diagonal or not.
We therefore \textit{assume} that $\tilde{b}(A)$ is diagonal.
Actually, this holds for all the examples discussed later.
Then the above equation reads
\beq
  \tilde{b}_\alpha(A) \delta_{\alpha \beta} = 
  \sum_{\lambda, \mu \in \E} 
  \components{\psi}{\alpha}{\lambda} A_{\mu \lambda}
  \components{\bar{\psi}}{\beta}{\mu} 
  = \sum_{\mu \in \E}
  \components{\psi}{\alpha}{A \mu} \components{\bar{\psi}}{\beta}{\mu} ,
\eeq
or equivalently,
\begin{subequations}
\label{bpsi}
\bea
  \tilde{b}(A) \psi &= \psi A^T ,  \\
  \tilde{b}_\alpha(A) \components{\psi}{\alpha}{\lambda} &=
  \components{\psi}{\alpha}{A \lambda} , \quad A \in \auto(\E) .
\eea
\end{subequations}

From the equations \eqref{Apsi} and \eqref{bpsi},
we obtain the transformation properties of the Cardy states
\begin{subequations}
\label{brane_charge}
\bea
  b(A) \ket{\alpha} &= \sum_{\lambda \in \E} 
    \components{\psi}{\alpha}{\lambda} b_\lambda(A) \dket{\lambda}
= \sum_{\lambda \in \E} 
\components{\psi}{A \alpha}{\lambda} \dket{\lambda}
= \ket{A \alpha} , \quad A \in \auto(\V) , \\
  A \ket{\alpha} &= \sum_{\lambda \in \E}
  \components{\psi}{\alpha}{\lambda} A \dket{\lambda}
  = \sum_{\lambda, \mu \in \E}
  \components{\psi}{\alpha}{\lambda} \dket{\mu} A_{\mu \lambda}
  = \sum_{\lambda, \mu \in \E}
  \components{\psi}{\alpha}{\lambda} \dket{\mu} 
  \delta_{A\mu, \lambda} \notag \\
  & = \sum_{\mu \in \E}
  \components{\psi}{\alpha}{A \mu} \dket{\mu}
  = \sum_{\mu \in \E} \tilde{b}_\alpha(A) 
  \components{\psi}{\alpha}{\mu} \dket{\mu}
  = \tilde{b}_\alpha(A) \ket{\alpha} , \quad A \in \auto(\E) .
\eea
\end{subequations}
The center $b(A) \in Z(G)$ induces a permutation of the Cardy states,
which is an automorphism of $\V$. 
On the other hand, $A \in \auto(\E)$ measures 
the `charge' (or the conjugacy class) of the Cardy states.

For a NIM-rep $n_\lambda$ of the fusion algebra, there corresponds a graph
whose vertices are labelled by the set $\V$ \cite{DZ,BPPZ}.
We can identify the boundary states with the vertices of the graph.
Then the automorphism group $\auto(\V)$ is naturally interpreted 
as the automorphism of the graph,
while $\tilde{b}_\alpha$ represents a coloring of the graph.

%%%%%%%%%%%%%%%%%%%%%%%%%%%%%%%%%%%%%%%%%%%%%%%%%%%%%%%%
\subsection{Untwisted states}

For the untwisted states, $\V = \E = \Spec(G)$ and $\psi = S$.
The equations \eqref{Apsi} and \eqref{bpsi} reduce to eq.\eqref{Ab}.
The transformation property \eqref{brane_charge} of the Cardy states
therefore reads
\beq
  b(A) \ket{\alpha} = \ket{A \alpha} , \quad
  A \ket{\alpha} = b_\alpha(A) \ket{\alpha} , \quad A \in \auto(g) .
\eeq

%%%%%%%%%%%%%%%%%%%%%%%%%%%%%%%%%%%%%%%%%%%%%%%%%%%%%%%%
\subsection{Twisted states}

For the twisted states, we have seen
\bes
  \E &= P_+^{k, \omega}(g^{(1)}) \simeq P_+^k(\tilde{g}^{(r)}) , \quad
  \V = P_+^k(g^{(r)}) , \\
  \components{\psi}{\alpha}{\lambda} &= \tilde{S}_{\alpha \tilde{\lambda}} .
\ees
From this form, it is natural to expect that
\beq
\label{twisted_auto}
  \auto(\E) \simeq \auto(\tilde{g}^{(r)}) , \quad
  \auto(\V) \simeq \auto(g^{(r)}) .
\eeq
We will show this is actually the case. We restrict ourselves to the case
of $g = A_{2l-1}$. 
The other $g$'s can be treated in the same way.

For $g = A_{2l-1}$, $g^{(r)} = A_{2l-1}^{(2)}$ and
$\tilde{g}^{(r)} = D_{l+1}^{(2)}$.
The explicit form of $\E$ and $\V$ reads
\bes
\label{A2l-1}
  \E &= P_+^{k, \omega}(A_{2l-1}^{(1)}) \\ &=
  \{(\lambda_1,\lambda_2,\cdots,\lambda_l,\cdots,\lambda_2,\lambda_1)\,|\,
  2\lambda_1 + 2\lambda_2 + \cdots + 2\lambda_{l-1} + \lambda_l \le k, \,
  \lambda_i \in \Z_{\ge 0} \} \\
  & \simeq 
  P_+^k(D_{l+1}^{(2)}) = 
  \{(\tilde{\lambda}_1,\tilde{\lambda}_2,\cdots,\tilde{\lambda}_l)\,|\,
  2\tilde{\lambda}_1 + 2\tilde{\lambda}_2 + \cdots + 
  2\tilde{\lambda}_{l-1} + \tilde{\lambda}_l \le k, \,
  \tilde{\lambda}_i \in \Z_{\ge 0} \}, \\
  \V &= P_+^k(A_{2l-1}^{(2)}) =
  \{(\alpha_1,\alpha_2,\cdots,\alpha_l) \,|\,
  \alpha_1 + 2\alpha_2 + \cdots + 2\alpha_l \le k, \,
  \alpha_i \in \Z_{\ge 0} \} .
\ees
First, note that
\beq
  \auto(A_{2l-1}^{(1)}) = \{1, A, A^2, \cdots, A^{2l-1}\} \simeq \Z_{2l}, 
\eeq
where the generator $A$ acts on $\lambda$ as
\bes
  A : (\lambda_1,\cdots,\lambda_{2l-2},\lambda_{2l-1}) \mapsto &
  (\lambda_0,\lambda_1,\cdots,\lambda_{2l-2}) , \\
  &\lambda_0 = k - (\lambda_1 + \cdots + \lambda_{2l-1}) .
\ees
Clearly, the elements that leave $\E$ invariant
are only $1$ and $A^l$, hence
\beq
  \auto(\E) = \{1, A^l \} \simeq \Z_2 .
\eeq
From eq.\eqref{A2l-1}, 
one can see that $A^l$ induces the following action on $P_+^k(D_{l+1}^{(2)})$
\bes
  A^l : (\tilde{\lambda}_1,\tilde{\lambda}_2,\cdots,
  \tilde{\lambda}_{l-1},\tilde{\lambda}_{l}) \mapsto &
  (\tilde{\lambda}_{l-1},\tilde{\lambda}_{l-2},\cdots,
  \tilde{\lambda}_{1},\tilde{\lambda}_{0}) , \\
  & \tilde{\lambda}_0 =
  k - (
  2\tilde{\lambda}_1 + \cdots +
  2\tilde{\lambda}_{l-1}+\tilde{\lambda}_{l}) , 
\ees
which is exactly the same as the action of the outer automorphism group
$\auto(D_{l+1}^{(2)}) \simeq \Z_2$. 
Hence, we have verified the first equality of eq.\eqref{twisted_auto},
$\auto(\E) \simeq \auto(\tilde{g}^{(r)})$.

The center of $SU(2l)$ is also $\Z_{2l}$, which is generated by $b(A)$.
The eigenvalue $b_\lambda(A)$ of $b(A)$ on the representation $\lambda$
reads
\beq
  b_\lambda(A) = 
  \exp\left(-\frac{\pi i}{l}((2l-1)\lambda_1 + (2l-2)\lambda_2 + \cdots
  + \lambda_{2l-1})\right) .
\eeq
On $\E$, this can be written as
\beq
  b_\lambda(A) = 
  \exp\left(-\frac{\pi i}{l}(2l \lambda_1 + 2l \lambda_2 + \cdots
  + 2l \lambda_{l-1} + l \lambda_l)\right) 
  = (-1)^{\lambda_l} .
\eeq
Hence, the stabilizer $S(\V)$ consists of $\{1,A^2,\cdots,A^{2l-2}\}$
and $S(\V) \simeq \Z_l$. 
The automorphism group $\auto(\V)$ is therefore
$\auto(\V) = \Z_{2l} / \Z_l = \{1, A \} \simeq \Z_2$. 
One can identify this with the outer automorphism group of $A_{2l-1}^{(2)}$,
which acts on $P_+^k(A_{2l-1}^{(2)})$ as
\bes
  A : (\alpha_1,\alpha_2,\cdots,\alpha_{l-1},\alpha_{l}) \mapsto &
  (\alpha_0,\alpha_2,\cdots,\alpha_{l-1},\alpha_{l}) , \\
  & \alpha_0 = k - (\alpha_1 + 2\alpha_2 + \cdots + 2\alpha_l) .
\ees
Actually, from the formula (proved in Appendix)
\begin{subequations}
\label{twisted_Ab}
\bea
  \tilde{S}_{A\alpha, \tilde{\lambda}} &=
  \tilde{S}_{\alpha \tilde{\lambda}} (-1)^{\tilde{\lambda}_l} , 
  \quad A \in \auto(A_{2l-1}^{(2)}) , \\
  \tilde{S}_{\alpha, \tilde{A}\tilde{\lambda}} &=
  (-1)^{\alpha_1 + 2\alpha_2 + \cdots + l \alpha_l}
  \tilde{S}_{\alpha \tilde{\lambda}} , 
  \quad \tilde{A} \in \auto(D_{l+1}^{(2)}) ,
\eea
\end{subequations}
one can confirm that the action of $\auto(\V)$ coincides with 
that of $\auto(A_{2l-1}^{(2)})$. Hence, we have verified the second equality
of eq.\eqref{twisted_auto}, $\auto(\V) \simeq \auto(g^{(r)})$.
Note that
the above formula \eqref{twisted_Ab} also exhibits that $\tilde{b}(A)$ in
eq.\eqref{bpsi} is diagonal, namely,
$\tilde{b}_\alpha(A) = (-1)^{\alpha_1 + 2\alpha_2 + \cdots + l \alpha_l}$.

%%%%%%%%%%%%%%%%%%%%%%%%%%%%%%%%%%%%%%%%%%%%%%%%%%%%%%%%
\subsection{$u(1)_k$}

The modular transformation matrix of the $u(1)_k$ theory \eqref{S_U1}
has the following symmetry
\beq
\label{Ab_U1}
  S_{m+1, m'} = S_{mm'} e^{-\frac{\pi i}{k} m'} .
\eeq
This is reminiscent of the relation \eqref{Ab}, and we regard 
$A : m \mapsto m+1$ as the generator of the automorphism group
\beq
  \auto(u(1)_k) = \{1, A, \cdots, A^{2k-1}\} \simeq \Z_{2k} .
\eeq
The `center' is defined in the same way
\beq
  b_m(A) = e^{-\frac{\pi i}{k} m} .
\eeq

Let us consider the twisted Cardy states
\beq
\label{u1_twisted}
  \ket{\pm} = 
  \frac{1}{\sqrt{2}} (\dket{0; \omega_c} \pm \dket{k; \omega_c}) . 
\eeq
The automorphism group $\auto(\E)$ is clearly $\{1, A^k\} \simeq \Z_2$,
under which $\ket{\pm}$ transforms as
\beq
  A^k \ket{\pm} = 
  \frac{1}{\sqrt{2}} (\dket{k; \omega_c} \pm \dket{0; \omega_c})
  = \pm \ket{\pm} .
\eeq
Since $b_k(A) = -1$, 
the stabilizer $S(\V)$ consists of $\{1,A^2,\cdots,A^{2k-2}\}$.
Hence, $\auto(\V) = \Z_{2k}/\Z_k = \{1, A\} \simeq \Z_2$, which acts
on $\ket{\pm}$ as
\beq
  b(A) \ket{\pm} = 
  \frac{1}{\sqrt{2}} (\dket{0; \omega_c} \mp \dket{k; \omega_c})
  = \ket{\mp} .
\eeq

%%%%%%%%%%%%%%%%%%%%%%%%%%%%%%%%%%%%%%%%%%%%%%%%%%%%%%%%
%%   Coset theories
%%%%%%%%%%%%%%%%%%%%%%%%%%%%%%%%%%%%%%%%%%%%%%%%%%%%%%%%
\section{Twisted boundary states in coset theories}

\subsection{Preliminaries}

Corresponding to the algebra embedding $h \subset g$, 
a representation $\lambda$ of $g$ is decomposed in terms of
the representations of $h$ as follows
\beq
\label{branching}
  (\lambda) \mapsto \oplus_{\mu} (\lambda; \mu) \otimes (\mu), \quad
  \lambda \in \Spec(G), \mu \in \Spec(H) .
\eeq
The spectrum of the $G/H$ coset theory is composed of all the possible
combination $(\lambda; \mu)$
\beq
\label{coset_spec}
  \Spec(G/H) = \{(\lambda; \mu) \,|\,
       \lambda \in \Spec(G), \mu \in \Spec(H), b_\lambda = b_\mu \}/
   (A \lambda; A \mu) \sim (\lambda; \mu) .
\eeq
Here the relation 
$(A \lambda; A \mu) \sim (\lambda; \mu), A \in \auto(h)$ is the
field identification,
\footnote{%
We do not consider the maverick cosets \cite{DJ,FSS2},
for which additional field identifications are necessary.
}%
and $b_\lambda = b_\mu$ is the selection rule for the common center of
$G$ and $H$.
To be precise, we should write the projected weight as $P\lambda$,
instead of $\lambda$, using the projection matrix $P$.
For simplicity, we omit this $P$, since it is obvious from the context 
whether $P$ should be appended or not.
In this paper, we restrict ourselves to the case that
all the identification orbit have the same length $N_0$
\beq
  N_0 = \lvert \{(A \lambda; A \mu)\,|\, A \in \auto(h)\}\rvert
      = \lvert \auto(h) \rvert .
\eeq
In particular, there is no fixed point in the field identification
\beq
  (A \lambda; A \mu) \neq (\lambda; \mu), \quad \text{for any } 
  \lambda \in \Spec(G), \mu \in \Spec(H), A \in \auto(h).
\eeq

The character of the coset theory is the branching function 
$\chi_{(\lambda; \mu)}$ of the algebra embedding $h \subset g$.
From the branching rule \eqref{branching}, we obtain
\beq
  \chi^G_\lambda = \sum_{\mu, b_\lambda = b_\mu} 
  \chi_{(\lambda; \mu)}\, \chi^H_\mu .
\eeq
The modular transformation of the coset characters can be written as
\bes
\label{coset_S}
  \chi_{(\lambda;\mu)}(-1/\tau) &= 
    \sum_{(\lambda';\mu') \in \Spec(G/H)} 
S_{(\lambda;\mu)(\lambda';\mu')}\, \chi_{(\lambda';\mu')}(\tau), \\
  S_{(\lambda;\mu)(\lambda';\mu')} &=
  N_0 S^G_{\lambda \lambda'} \bar{S}^H_{\mu \mu'} .
\ees
This $S$-matrix has several properties necessary for a consistent
theory.
First, $S_{(\lambda;\mu)(\lambda';\mu')}$ does not depend on the representative 
of the field identification orbit. Namely,
\beq
  S_{(A\lambda;A\mu)(\lambda';\mu')} 
  = N_0 S^G_{A\lambda, \lambda'} \bar{S}^H_{A\mu, \mu'}
  = N_0 S^G_{\lambda \lambda'} \bar{S}^H_{\mu \mu'} 
       b_{\lambda'}(A) b_{\mu'}(A)^{-1}
  = S_{(\lambda;\mu)(\lambda';\mu')} .
\eeq
Here we used the property \eqref{Ab} for $S^G$ and $S^H$. 
The last equality follows from the selection rule $b_{\lambda'} = b_{\mu'}$. 
Next, most importantly, $S_{(\lambda;\mu)(\lambda';\mu')}$ is unitary
\bes
  \sum_{(\lambda';\mu') \in \Spec(G/H)}
  S_{(\lambda;\mu)(\lambda';\mu')} \bar{S}_{(\lambda'';\mu'')(\lambda';\mu')} 
  &= \sum_{(\lambda';\mu')} N_0^2 
     S^G_{\lambda \lambda'} \bar{S}^H_{\mu \mu'}
 \bar{S}^G_{\lambda'' \lambda'} S^H_{\mu'' \mu'} \\
  &= \frac{1}{N_0} \sum_{\lambda', \mu'} 
     \frac{1}{N_0} \sum_{A \in \auto(h)} b_{\lambda'}(A) b_{\mu'}(A)^{-1} \\
 & \quad\quad\quad\quad\quad\quad\quad \times N_0^2
 S^G_{\lambda \lambda'} \bar{S}^G_{\lambda'' \lambda'}
 \bar{S}^H_{\mu \mu'} S^H_{\mu'' \mu'} \\
  &= \sum_{A \in \auto(h)} \sum_{\lambda', \mu'} 
     S^G_{A\lambda, \lambda'} \bar{S}^G_{\lambda'' \lambda'}
 \bar{S}^H_{A\mu, \mu'} S^H_{\mu'' \mu'} \\
  &= \sum_{A \in \auto(h)} 
     \delta_{A\lambda, \lambda''} \delta_{A\mu, \mu''} \\
  &= \delta_{(\lambda;\mu)(\lambda'';\mu'')} .
\ees
Here we used our assumption of no fixed points to rewrite the sum
\beq
  \sum_{(\lambda';\mu') \in \Spec(G/H)} \rightarrow \quad
     \frac{1}{N_0} \sum_{\lambda', \mu'} 
   \frac{1}{N_0} \sum_{A \in \auto(h)} b_{\lambda'}(A) b_{\mu'}(A)^{-1} .
\eeq
The projection operator introduced above takes account of the selection
rule. We parametrized the center of $H$ by the elements $A$ of $\auto(h)$
using the isomorphism $Z(H) \simeq \auto(h)$. 

The fusion algebra of the coset theory can be obtained via the Verlinde 
formula \eqref{Verlinde}
\begin{subequations}
\label{coset_fusion}
\bes
  \fusion{(\lambda;\mu)}{(\lambda';\mu')}{(\lambda'';\mu'')}
  &= \sum_{(\rho;\sigma) \in \Spec(G/H)} 
    \frac{%
  S_{(\lambda;\mu)(\rho;\sigma)}
  S_{(\lambda';\mu')(\rho;\sigma)} 
  \bar{S}_{(\lambda'';\mu'')(\rho;\sigma)}}
  {S_{(0,0)(\rho;\sigma)}}  \\
  &= \frac{1}{N_0} \sum_{\rho, \sigma} 
 \frac{1}{N_0} \sum_{A \in \auto(h)} b_{\rho}(A) b_{\sigma}(A)^{-1} \\
 & \quad\quad\quad\quad\quad\quad\quad\quad\quad
 \times N_0^2 
 \frac{%
 S^G_{\lambda \rho}S^G_{\lambda' \rho}\bar{S}^G_{\lambda'' \rho}}
 {S^G_{0 \rho}}
 \frac{%
 \bar{S}^H_{\mu \sigma}\bar{S}^H_{\mu' \sigma} S^H_{\mu'' \sigma}}
  {\bar{S}^H_{0 \sigma}} \\
  &= \sum_{A \in \auto(h)} 
     \mathcal{N}^G_{A\lambda \,\lambda'}{}^{\lambda''}
 \mathcal{N}^H_{A\mu\,\mu'}{}^{\mu''} .
\ees
In the matrix form,
\beq
  N_{(\lambda;\mu)} = \sum_{A \in \auto(h)} N^G_{A\lambda} \otimes N^H_{A\mu},
\eeq
\end{subequations}
where both the columns and the rows are restricted to $\Spec(G/H)$. 

%%%%%%%%%%%%%%%%%%%%%%%%%%%%%%%%%%%%%%%%%%%%%%%%%%%%%%%%
\subsection{Boundary states}

The boundary condition of the coset theory $G/H$ follows from that of the current
algebras $G$ and $H$. 
If we adopt the untwisted boundary condition \eqref{symmetric_bc} for $G$ and
$H$, we have the untwisted boundary condition for $G/H$. 
The resulting boundary states are the untwisted Cardy states
\beq
  \ket{(\lambda; \mu)} = \sum_{(\lambda'; \mu') \in \Spec(G/H)}
  S_{(\lambda;\mu)(\lambda';\mu')} \dket{(\lambda'; \mu')} ,
\eeq
where $\dket{(\lambda; \mu)}$ is the Ishibashi state for
the primary field $(\lambda; \mu)$ normalized in the same way as before
\beq
  \dbra{(\lambda; \mu)} \tilde{q}^{H_c} \dket{(\lambda; \mu)}
  = \frac{1}{S_{(0;0)(\lambda;\mu)}} \chi_{(\lambda;\mu)}(-1/\tau) .
\eeq
Suppose that
the current algebra $G$ admits an automorphism $\omega$ of the horizontal
algebra and that $\omega$ induces an automorphism of $H$.
Then both the current algebras $G$ and $H$ can be twisted and we have
the twisted boundary condition of the coset theory $G/H$. 
We have seen in the previous sections that 
there exists a NIM-rep of the fusion algebra for each set of
the mutually consistent boundary states satisfying the Cardy condition.
The regular NIM-rep $N_\lambda$ corresponds to the untwisted boundary states,
while for the twisted states we have a non-trivial NIM-rep. 
We should find a non-trivial NIM-rep of the fusion algebra
\eqref{coset_fusion} for the twisted Cardy states in the coset theory.

Finding a NIM-rep is nothing but finding a diagonalization matrix 
$\psi$ of the fusion algebra. 
For the regular NIM-rep, $\psi$ coincides with the modular transformation
matrix $S$, which is related with those of $G$ and $H$ as follows
\beq
\label{coset_S2}
  S_{(\lambda;\mu)(\lambda';\mu')} =
  N_0 S^G_{\lambda \lambda'} \bar{S}^H_{\mu \mu'} .
\eeq
This form suggests the following expression for the twisted boundary states
in the coset theory
\beq
\label{coset_psi}
  \components{\psi}{(\alpha; \beta)}{(\lambda;\mu)} =
  N \psi^G_{\alpha}{}^{\lambda} \bar{\psi}^H_{\beta}{}^{\mu} ,
\eeq
where $\psi^G$ and $\psi^H$ are the boundary state coefficients
for the twisted Cardy states in $G$ and $H$, respectively, 
and $N$ is an integer that divides $N_0$. 
We shall show that this actually realizes a NIM-rep of the fusion algebra
\eqref{coset_fusion} of the coset theory. 

The label $(\alpha; \beta)$ of the boundary states
is composed of those of the current algebra theories. 
However, not all the combination of $\alpha \in \V^G$ and $\beta \in \V^H$
is allowed, since $(\lambda; \mu)$ should belong to the spectrum
\eqref{coset_spec} of the coset theory.

First, from the field identification
$(A\lambda; A\mu) \sim (\lambda; \mu), \, A \in \auto(h),$
it should be satisfied that
\beq
\label{coset_A}
  \components{\psi}{(\alpha; \beta)}{(A\lambda; A\mu)} =
  \components{\psi}{(\alpha; \beta)}{(\lambda; \mu)} , \quad
  A \in \auto(\E^H) .
\eeq
Here we restrict the identification group of the spectrum to $\auto(\E^H)$
since $\psi^H$ is defined only in $\E^H$. 
Correspondingly, the length of the identification orbit
is shorter than $N_0$. We set the integer $N$ in \eqref{coset_psi}
to the length of this orbit
\beq
\label{def_N}
  N = |\{(A \lambda; A \mu)\,|\, A \in \auto(\E^H) \}| .
\eeq
The requirement \eqref{coset_A} implies that
\bes
  \components{\psi}{(\alpha; \beta)}{(\lambda; \mu)} =
  \components{\psi}{(\alpha; \beta)}{(A \lambda;A \mu)} &=
  N \psi^G_{\alpha}{}^{A\lambda} \bar{\psi}^H_{\beta}{}^{A\mu} \\
  &= N \psi^G_{\alpha}{}^{\lambda} \bar{\psi}^H_{\beta}{}^{\mu}
    \tilde{b}_\alpha(A) \tilde{b}_\beta(A)^{-1} \\
  &= \components{\psi}{(\alpha; \beta)}{(\lambda;\mu)}
    \tilde{b}_\alpha(A) \tilde{b}_\beta(A)^{-1} ,
\ees
where we used the relation \eqref{bpsi} for $\psi^G$ and $\psi^H$.
Hence we have to require $\tilde{b}_\alpha = \tilde{b}_\beta$ for
$(\alpha; \beta)$ to be a label of the boundary state. 

Next, from the selection rule $b_\lambda = b_\mu$ for $(\lambda; \mu)$,
we obtain
\bes
  \components{\psi}{(\alpha; \beta)}{(\lambda;\mu)} =
  \components{\psi}{(\alpha; \beta)}{(\lambda;\mu)}
  b_\lambda(A) b_\mu(A)^{-1} &=
  N \psi^G_{A\alpha}{}^{\lambda} \bar{\psi}^H_{A\beta}{}^{\mu} \\
  &= \components{\psi}{(A\alpha; A\beta)}{(\lambda;\mu)} ,
\ees
where we used the relation \eqref{Apsi}.
Therefore we should identify
$(A\alpha; A\beta), \, A \in \auto(\V^H),$ with $(\alpha; \beta)$.
Together with the above result, 
we define the set $\V^{G/H}$ of the labels of the twisted boundary states
in the coset theory as follows
\beq
\label{coset_V}
  \V^{G/H} = \{(\alpha; \beta)\,|\,
       \alpha \in \V^G, \beta \in \V^H, \tilde{b}_\alpha = \tilde{b}_\beta \}/
   (A \alpha; A \beta) \sim (\alpha; \beta) .
\eeq
We can see the structure of $\V^{G/H}$ is exactly parallel to that of
$\Spec(G/H)$. Namely, we have the counterpart of the field identification and
the selection rule in $\V^{G/H}$. 
We call the identification $(A \alpha; A \beta) \sim (\alpha; \beta)$
and the selection rule $\tilde{b}_\alpha = \tilde{b}_\beta$ for 
the boundary states
the brane identification and the brane selection rule, respectively.

With this definition at hand, we can check that
the boundary state coefficients \eqref{coset_psi} actually 
realizes a NIM-rep of the fusion algebra.
In order to do that, we restrict ourselves to the case
that the length of the identification orbit
$\{(A \alpha; A\beta)\,|\, A \in \auto(\V^H)\}$ is also given by $N$ 
\eqref{def_N}.
This is not so restrictive and holds for all the examples discussed below.
Let us first check the unitarity of
$\components{\psi}{(\alpha;\beta)}{(\lambda;\mu)}$
\bes
  \sum_{(\lambda;\mu) \in \E^{G/H}}
  \components{\psi}{(\alpha;\beta)}{(\lambda;\mu)}
  \components{\bar{\psi}}{(\alpha';\beta')}{(\lambda;\mu)}
  &= \sum_{(\lambda;\mu)} N^2 
     \psi^G_\alpha{}^\lambda \bar{\psi}^H_\beta{}^\mu
 \bar{\psi}^G_{\alpha'}{}^\lambda \psi^H_{\beta'}{}^\mu \\
  &= \frac{1}{N} \sum_{\lambda, \mu} 
     \frac{1}{N} \sum_{A \in \auto(\V^H)} b_{\lambda}(A) b_{\mu}(A)^{-1} \\
 & \quad\quad\quad\quad\quad\quad\quad \times N^2
 \psi^G_\alpha{}^\lambda \bar{\psi}^G_{\alpha'}{}^\lambda
 \bar{\psi}^H_\beta{}^\mu \psi^H_{\beta'}{}^\mu \\
  &= \sum_{A \in \auto(\V^H)} \sum_{\lambda, \mu} 
     \psi^G_{A\alpha}{}^\lambda \bar{\psi}^G_{\alpha'}{}^\lambda
 \bar{\psi}^H_{A\beta}{}^\mu \psi^H_{\beta'}{}^\mu \\
  &= \sum_{A \in \auto(\V^H)} 
     \delta_{A\alpha, \alpha'} \delta_{A\beta, \beta'} .
\ees
From this calculation, one can see that
$\components{\psi}{(\alpha;\beta)}{(\lambda;\mu)}$ is unitary
unless there are fixed points in the brane identification
$(A\alpha;A\beta) \sim (\alpha;\beta)$. 
If, for example, there is a fixed point
$A\alpha=\alpha, A\beta=\beta,$ for $A^2 = 1$,
we have the result
$\sum_{(\lambda;\mu)}
 \components{\psi}{(\alpha;\beta)}{(\lambda;\mu)}
 \components{\bar{\psi}}{(\alpha';\beta')}{(\lambda;\mu)} = 2$
instead of $1$. 
This is the situation familiar in the field identification of
the coset theory and we need some resolution of the fixed point
in order to have a consistent theory \cite{FSS2}.

The NIM-rep associated with
$\components{\psi}{(\alpha;\beta)}{(\lambda;\mu)}$ can be obtained
in the same way as above
\begin{subequations}
\label{coset_NIM}
\bes
  \fusion[n]{(\lambda;\mu)}{(\alpha;\beta)}{(\alpha';\beta')}
  &= \sum_{(\rho;\sigma) \in \E^{G/H}}
     \components{\psi}{(\alpha;\beta)}{(\rho;\sigma)}
    \frac{S_{(\lambda;\mu)(\rho;\sigma)}} 
  {S_{(0,0)(\rho;\sigma)}}
  \components{\bar{\psi}}{(\alpha';\beta')}{(\rho;\sigma)} \\
  &= \frac{1}{N} \sum_{\rho, \sigma} 
 \frac{1}{N} \sum_{A \in \auto(\V^H)} b_{\rho}(A) b_{\sigma}(A)^{-1} \\
 & \quad\quad\quad\quad\quad\quad\quad\quad\quad
 \times N^2
 \psi^G_{\alpha}{}^\rho
 \frac{S^G_{\lambda \rho}}{S^G_{0 \rho}}
 \bar{\psi}^G_{\alpha'}{}^\rho \,
 \bar{\psi}^H_{\beta}{}^\sigma
 \frac{\bar{S}^H_{\mu \sigma}}{\bar{S}^H_{0 \sigma}}
 \psi^H_{\beta'}{}^\sigma \\
  &= \sum_{A \in \auto(\V^H)} 
     n^G_{A\lambda \,\alpha}{}^{\alpha'}
 n^H_{A\mu\,\beta}{}^{\beta'} .
\ees
In the matrix form,
\beq
  n_{(\lambda;\mu)} =
  \sum_{A \in \auto(\V^H)} n^G_{A\lambda} \otimes n^H_{A\mu},
\eeq
\end{subequations}
where both the columns and the rows are restricted to $\V^{G/H}$. 
Since the components of $n^G$ and $n^H$ are non-negative integers,
$n_{(\lambda;\mu)}$ is also non-negative integer matrix.
Hence $n_{(\lambda;\mu)}$ is a NIM-rep of the fusion algebra,
if there is no fixed point in the brane identification.

%%%%%%%%%%%%%%%%%%%%%%%%%%%%%%%%%%%%%%%%%%%%%%%%%%%%%%%%
%%   Examples
%%%%%%%%%%%%%%%%%%%%%%%%%%%%%%%%%%%%%%%%%%%%%%%%%%%%%%%%
\section{Examples}

In this section, we apply the methods developed in the previous section
to obtain the twisted boundary states in various coset theories.

\subsection{$su(3)_k \oplus su(3)_l/su(3)_{k+l}$}
The diagonal coset of $su(3)$ is the simplest example that
admits the twisted boundary conditions. 
We consider the case of $su(3)_1 \oplus su(3)_1/su(3)_2$, 
although our methods can be applied to the other levels. 

Let us start with the twisted boundary states in the $su(3)$ theory
at level $k$.
The automorphism $\omega$ of the horizontal subalgbra
acts on the weight of $su(3)$ as follows
\beq
  \omega : (\lambda_1, \lambda_2) \mapsto (\lambda_2, \lambda_1) .
\eeq
The spectrum $\E$ invariant under $\omega$ reads
\beq
  \E = P_+^{k,\omega}(A_2^{(1)})
    = \{(\lambda_1, \lambda_1)\,|\,
        2\lambda_1 \le k, \, \lambda_1 \in \Z_{\ge 0} \} . 
\eeq
The boundary states are labelled by the integrable representation $\alpha$
of the twisted affine Lie algebra $A_2^{(2)}$
\beq
  \V = P_+^k(A_2^{(2)})
    = \{\alpha = (\alpha_1)\,|\,
        2 \alpha_1 \le k, \, \alpha \in \Z_{\ge 0} \}, 
\eeq
and take the form
\beq
  \ket{\alpha; \omega} = \sum_{\lambda \in \E}
  \tilde{S}_{\alpha \tilde{\lambda}} \dket{\lambda; \omega} .
\eeq
Here $\tilde{S}$ is the modular transformation matrix of $A_2^{(2)}$
\beq
  \tilde{S}_{\lambda \mu} = \frac{2}{\sqrt{k+3}} 
  \sin\left(\frac{2\pi}{k+3}(\lambda_1 + 1)(\mu_1 + 1)\right) , \quad
  \lambda,  \mu \in P_+^{k}(A_2^{(2)}) ,
\eeq
and $\tilde{\lambda}$ is defined as $\tilde{\lambda}_1 = \lambda_1$.

For $k=1$, 
$\E = \{(0,0)\}$ and
$\V = P_+^{k=1}(A_2^{(2)}) = \{(0)\}$.
Hence there exists only one twisted state
\beq
 \ket{0; \omega} = \tilde{S}_{00} \dket{(0,0); \omega} 
 = \dket{(0,0); \omega} .
\eeq
For $k=2$, $\E = \{(0,0),(1,1)\}$ and 
$\V = P_+^{k=2}(A_2^{(2)}) = \{(0), (1)\}$.
We have two twisted states
\beq
  \ket{\alpha; \omega} = \sum_{\lambda = 0, 1}
  \tilde{S}_{\alpha \lambda} \dket{(\lambda, \lambda); \omega} , \quad
  \alpha = 0, 1, 
\eeq
where $\tilde{S}$ takes the form
\beq
  \tilde{S} = \frac{2}{\sqrt{5}}
  \begin{pmatrix} \sin \frac{2\pi}{5} & \sin \frac{\pi}{5} \\
                  \sin \frac{\pi}{5} & -\sin \frac{2\pi}{5} \end{pmatrix} .
\eeq

Since the diagonal action of $\omega$ on $su(3)_1 \oplus su(3)_1$
induces the automorphism of $su(3)_2 \subset su(3)_1 \oplus su(3)_1$,
we have the twisted boundary condition in the coset theory
$su(3)_1 \oplus su(3)_1/su(3)_2$.
Since $A_{2}^{(2)}$ has no outer automorphism, 
$\auto(\E) = \auto(\V) = \{1\}$. 
Hence both the brane identification and the brane selection rule
is trivial.
We therefore obtain two twisted boundary states in the coset theory
\bes
\label{A2}
  \ket{(0,0;0); \omega} &=
  \frac{2}{\sqrt{5}} \left(\sin \frac{2\pi}{5} \dket{(0,0;(0,0)); \omega}
  + \sin \frac{\pi}{5} \dket{(0,0;(1,1)); \omega}\right) ,\\
  \ket{(0,0;1); \omega} &=
  \frac{2}{\sqrt{5}} \left(\sin \frac{\pi}{5} \dket{(0,0;(0,0)); \omega}
  - \sin \frac{2\pi}{5} \dket{(0,0;(1,1)); \omega}\right) .
\ees

%%%%%%%%%%%%%%%%%%%%%%%%%%%%%%%%%%%%%%%%%%%%%%%%%%%%%%%%
\subsection{$su(4)_k \oplus su(4)_l/su(4)_{k+l}$}

The diagonal coset of $su(4) = A_3$ can be treated in the same way
as $su(3)$. 

The automorphism $\omega$ acts on the weight of $su(4)$ as
\beq
  \omega : (\lambda_1,\lambda_2,\lambda_3) \mapsto
     (\lambda_3,\lambda_2,\lambda_1) .
\eeq
Hence the spectrum $\E$ reads
\beq
  \E = P_+^{k,\omega}(A_3^{(1)})
    = \{(\lambda_1, \lambda_2, \lambda_1)\,|\, 2\lambda_1 + \lambda_2 \le k,
\, \lambda_i \in \Z_{\ge 0} \} . 
\eeq
The boundary states are labelled by the integrable representation $\alpha$
of the twisted affine Lie algebra $A_3^{(2)}$
\beq
  \V = P_+^k(A_3^{(2)})
    = \{\alpha = (\alpha_1, \alpha_2)\,|\,
\alpha_1 + 2\alpha_2 \le k, \, \alpha \in \Z_{\ge 0} \}, 
\eeq
and take the form
\beq
  \ket{\alpha; \omega} = \sum_{\lambda \in \E}
  \tilde{S}_{\alpha \tilde{\lambda}} \dket{\lambda; \omega} .
\eeq
Here $\tilde{S}$ is the modular transformation matrix between
$A_3^{(2)}$ and $D_3^{(2)} \simeq A_3^{(2)}$. 

For $k=1$, $\E = \{(0,0,0), (0,1,0)\}$ and $\V = \{(0,0), (1,0)\}$.
Hence we have two twisted boundary states, for which the coefficients
read
\beq
  \tilde{S} = \frac{1}{\sqrt{2}}
  \begin{pmatrix} 1 & 1 \\
                  1 & -1 \end{pmatrix} .
\eeq
For $k=2$, $\E = \{(0,0,0),(0,2,0),(1,0,1),(0,1,0)\}$ and
$\V = \{(0,0),(2,0),(0,1),(1,0)\}$. 
We therefore have four twisted boundary states, for which the coefficients
read
\beq
\label{A3_k2}
  \tilde{S} = \frac{1}{\sqrt{6}}
  \begin{pmatrix} 1 & 1  & 1 & \sqrt{3}\\
                  1 & 1  & 1 & -\sqrt{3}\\
  1 & 1  & -2 & 0 \\
  \sqrt{3} & -\sqrt{3} & 0 & 0 \end{pmatrix} .
\eeq

Since $\auto(A_3^{(2)}) = \{1, A\} \simeq \Z_2$, 
we need the brane identification
in the coset theory $su(4)_1 \oplus su(4)_1/su(4)_2$. 
The length of the identification orbit is $2$.
The generator $A$ of $\auto(A_3^{(2)})$ acts on $\V$ as
\beq
  A : (\alpha_1,\alpha_2) \mapsto (\alpha_0, \alpha_2) , \quad
   \alpha_0 = k - (\alpha_1 + 2 \alpha_2) . 
\eeq
On the other hand, $\auto(D_3^{(2)}) = \{1, \tilde{A}\} \simeq \Z_2$
acts on $\E$ as
\beq
  \tilde{A} : (\lambda_1,\lambda_2,\lambda_1) \mapsto
      (\lambda_1,\lambda_0,\lambda_1) , \quad
  \lambda_0 = k -(2\lambda_1 + \lambda_2) .
\eeq
From the formula \eqref{twisted_Ab}, we obtain
\beq
  \tilde{b}_\alpha(\tilde{A}) = (-1)^{\alpha_1} .
\eeq
Putting these facts together, we can write down
the set $\V^{G/H}$ of the label of the twisted boundary states for
$su(4)_1 \oplus su(4)_1/su(4)_2$
\bes
  \V^{G/H} = & \{%
  ((0,0),(0,0);(0,0)),((0,0),(0,0);(2,0)), \\ 
  & \quad\quad\quad\quad ((0,0),(0,0);(0,1)), ((0,0),(1,0);(1,0)) \} .
\ees
The boundary state coefficients can be calculated
by the formula \eqref{coset_psi} and coincide with $\tilde{S}$ for $k=2$
\eqref{A3_k2}.

%%%%%%%%%%%%%%%%%%%%%%%%%%%%%%%%%%%%%%%%%%%%%%%%%%%%%%%%
\subsection{$su(2)_k/u(1)_k$}

The $su(2)_k/u(1)_k$ parafermion theory ($PF_k$) is the simplest example
including the $u(1)$ factor. 
This theory is equivalent with the $su(k)_1 \oplus su(k)_1/su(k)_2$ theory.
Therefore we can check the validity of our procedure by comparing
the result with that obtained above.

The spectrum of the parafermion theory reads
\bes
  &\Spec(PF_k) = \\
  &\quad \{(l;m)\,|\, l = 0,1,\cdots,k, m \in \Z/2k \Z, l = m \mod 2 \}/
  (k - l; m + k) \sim (l; m) ,
\ees
where $l \in P_+^k(A_1^{(1)})$ stands for the integrable representation
of $su(2)_k$ while $m \in \Spec(u(1)_k)$ is the irreducible representation
of $u(1)_k$.

Although the charge conjugation $\omega_c$ is an inner automorphism in $su(2)$,
it induces an outer automorphism $\omega_c$ of $u(1) \subset su(2)$.
Hence $\omega_c$ is an outer automorphism of the coset theory and
we obtain the boundary states twisted by $\omega_c$ \cite{MMS}.
Since $\omega_c$ is inner in $su(2)$, we use the untwisted boundary
states $\ket{l}$ for the $su(2)$ sector
\footnote{%
To be precise, we have to twist the boundary states by
the charge conjugation. 
We omit it since
it does not affect the coefficient
$S^{su(2)}$ of the boundary states.}%
\beq
  \ket{l} = \sum_{l'} S^{su(2)}_{ll'} \dket{l'} ,
\eeq
where the modular transformation matrix reads
\beq
  S^{su(2)}_{ll'} = 
  \sqrt{\frac{2}{k+2}} \sin \left(\frac{\pi}{k+2}(l+1)(l'+1)\right) .
\eeq 
The $u(1)$ part is described by the twisted boundary states $\ket{\pm}$
\eqref{u1_twisted}.

The brane identification and the brane selection rule are applied
in the same way as the previous examples.
However, there is a subtlety for $k \in 2\Z_{\ge 0}$. 
Since the center $b$ of $SU(2)$ acts on $m \in \Spec(u(1)_k)$ as $(-1)^m$,
the Ishibashi state $\dket{m=k; \omega_c}$ transforms as
$b \dket{m=k; \omega_c} = (-1)^k \dket{m=k; \omega_c}$.
For odd $k$, this induces the automorphism of the boundary states
$\ket{\pm} \rightarrow \ket{\mp}$.
For even $k$, however, the action of $b$ leaves $\ket{\pm}$ invariant and
the automorphism group $\auto(\V^{u(1)})$ is trivial
(the stabilizer $S(\V^{u(1)})$ coincides with $\auto(su(2)_k) \simeq \Z_2$).
The set $\V^{PF_k}$ of the label of the twisted boundary states 
in the parafermion theory therefore reads
\begin{subequations}
\bes
  \V^{PF_{\text{odd } k}} &= \{(l;(-1)^l)\,|\, l = 0,1,\cdots,k \}
  / (k-l; +) \sim (l; -) \\
  &= \{(l;(-1)^l)\,|\, l = 0,1, \cdots, (k-1)/2 \} \quad \text{for odd $k$}
\ees
and 
\bes
  \V^{PF_{\text{even } k}} &= \{(l;(-1)^l)\,|\, l = 0,1,\cdots,k \}
  / (k-l; \pm) \sim (l; \pm) \\
  &= \{(l;(-1)^l)\,|\, l = 0,1,\cdots,k/2 \} \quad \text{for even $k$}
\ees
The set $\V^{PF_k}$ appears to have the same structure irrespective
of the parity of $k$. 
However, for even $k$, $(l;(-1)^{l}) \sim (k-l;(-1)^{l})$ and
we have a fixed point $(k/2; (-1)^{k/2})$ 
in the brane identification.
In order to have a complete set of the boundary states, we have to
resolve the fixed point. This is possible because we have one 
additional Ishibashi state $\dket{(k/2;k/2);\omega_c}$ for even $k$
\cite{MMS} and we obtain the resolved set 
\beq
  \tilde{\V}^{PF_{\text{even } k}} = 
  \{(l;(-1)^l)\,|\, l = 0,1,\cdots,k/2 - 1 \} \cup \{(k/2;(-1)^{k/2})_\pm \} .
\eeq
\end{subequations}
The boundary state coefficients follow from the formula \eqref{coset_psi}.
For the ordinary states, we obtain
\begin{subequations}
\beq
  \ket{(l;(-1)^l);\omega_c} = \sum_{l' \text{ even}}
  \sqrt{2} S^{su(2)}_{ll'} \dket{(l';0); \omega_c}, \quad l < \frac{k}{2} ,
\eeq
while for the `fractional' states we have
\beq
  \ket{(k/2;(-1)^{k/2})_\pm;\omega_c} = \sum_{l' \text{ even}}
  \frac{1}{\sqrt{2}} S^{su(2)}_{ll'} \dket{(l';0); \omega_c}
  \pm \frac{1}{\sqrt{2}} \dket{(k/2;k/2); \omega_c} .
\eeq
\end{subequations}
This reproduces the result obtained in \cite{MMS}. 

Let us calculate the explicit form of the boundary states 
for $k = 3,4$, and compare it with the results for the diagonal coset.
For $k = 3$, 
$\E = \{(0;0),(2;0)\}$ and $\V = \{(0;+),(1;-)\}$.
In this basis, the boundary state coefficient $\psi$ reads
\beq
  \psi = \frac{2}{\sqrt{5}} \begin{pmatrix}
  \sin \frac{\pi}{5} & \sin \frac{2\pi}{5} \\
  \sin \frac{2\pi}{5} & -\sin \frac{\pi}{5} \end{pmatrix} .
\eeq
This coincides with the result \eqref{A2} for the $su(3)$ diagonal coset
after the identification
$\dket{(0,0;(1,1))} = -\dket{(2,0)}$,
$\ket{(0,0;0)} = \ket{(2;0)}$ and $\ket{(0,0;1)} = \ket{(0;0)}$.

For $k=4$,
$\E = \{(0;0), (2;0), (4;0), (2;2)\}$ and 
$\V = \{(0;+), (1;-), (2;+)_\pm \}$. 
The boundary state coefficient reads
\beq
  \psi = \frac{1}{\sqrt{6}} \begin{pmatrix}
  1 & 2 & 1 & 0 \\
  \sqrt{3} & 0 & -\sqrt{3} & 0 \\
  1 & -1 & 1 & \sqrt{3} \\
  1 & -1 & 1 & -\sqrt{3} \end{pmatrix} ,
\eeq
which again coincides with the result \eqref{A3_k2} for 
the $su(4)$ diagonal coset after an appropriate identification of the states.

%%%%%%%%%%%%%%%%%%%%%%%%%%%%%%%%%%%%%%%%%%%%%%%%%%%%%%%%
%%   Summary
%%%%%%%%%%%%%%%%%%%%%%%%%%%%%%%%%%%%%%%%%%%%%%%%%%%%%%%%
\section{Summary}

In this paper, we have developed the method for constructing
the twisted boundary states in the $G/H$ coset conformal field theory.
In the way analogous to the field identification and the selection rule
of the coset theory, we introduce the notion of the brane identification
and the brane selection rule which act on the set of the boundary states.
We have shown that the twisted boundary states of the $G/H$ theory 
follow from those of the $G$ and the $H$ theories making use of these rules.
As a check of our procedure, we have treated in detail
the $su(n)_1 \oplus su(n)_1/su(n)_2$ theory and the $su(2)_k/u(1)_k$ 
parafermion theory, which are equivalent with each other, and
have obtained the consistent results.
Also, we have seen that our boundary states for the parafermion theory 
reproduces the results obtained in \cite{MMS}.

In this paper, we have restricted ourselves to the charge-conjugation
(or the diagonal) modular invariant.
It is interesting to extend our analysis to
other non-trivial modular invariants.
The minimal models have the description as the coset theory,
namely, $su(2)_k \oplus su(2)_1/su(2)_{k+1}$. 
One can easily verify that our procedure, in particular
the formula \eqref{coset_psi}, yields all the boundary states 
of the minimal models obtained in \cite{BPPZ},
by appropriately extending the brane identification
and the selection rule to the $D$ and the $E$ type modular invariants.
The related problem is the issue of the unphysical NIM-reps.
We can formally construct a NIM-rep of the $G/H$ theory 
starting from an unphysical NIM-rep of the $G$ (or $H$) theory.
It is interesting to determine whether the resulting NIM-rep is
physical or not, although it is likely to be unphysical.

It is also interesting to apply our method to the super coset theories,
especially the Kazama-Suzuki models \cite{KS}.

We have seen there are fixed points in the brane identification,
and we should resolve them to obtain the consistent theory.
This phenomena is the brane version of the field identification fixed
points.
Hence, in order to have a deep understanding of this, 
it will be necessary to
extend our analysis to the case of the coset theory
with the field identification fixed points.

%%%%%%%%%%%%%%%%%%%%%%%%%%%%%%%%%%%%%%%%%%%%%%%%%%%%%%%%
\vskip \baselineskip
\noindent
\textbf{Acknowledgement: }
I would like to thank H.~Awata, M.~Kato, M.~Oshikawa and Y.~Satoh
for helpful discussions.

%%%%%%%%%%%%%%%%%%%%%%%%%%%%%%%%%%%%%%%%%%%%%%%%%%%%%%%%
%%   Appendix
%%%%%%%%%%%%%%%%%%%%%%%%%%%%%%%%%%%%%%%%%%%%%%%%%%%%%%%%
\newpage
\appendix
%%%%%%%%%%%%%%%%%%%%%%%%%%%%%%%%%%%%%%%%%%%%%%%%%%%%%%%%
\section{Modular transformation matrix of
$(A_{2l-1}^{(2)}, D_{l+1}^{(2)})$}

In this Appendix, we show the transformation property
of the modular transformation matrix between
$A_{2l-1}^{(2)}$ and $D_{l+1}^{(2)}$ under the action of
the outer automorphism group of the algebras.
The derivation is exactly parallel to the case of the untwisted
affine Lie algebras (see, for example, \S 14.6 of \cite{FMS}).

The modular transformation of the characters of
the twisted affine Lie algebra $A_{2l-1}^{(2)}$ gives rise to
those of $D_{l+1}^{(2)}$, and vice versa. 
The integrable representations of these algebras at level $k$
are labelled as follows
\bes
  P_+^k(A_{2l-1}^{(2)}) &= 
  \{(\lambda_1,\lambda_2, \cdots, \lambda_l)\,|\,
    \lambda_1 + 2 \lambda_2 + \cdots + 2\lambda_l \le k, 
\lambda_i \in \Z_{\ge 0} \}, \\
  P_+^k(D_{l+1}^{(2)}) &= 
  \{(\mu_1, \mu_2, \cdots, \mu_l)\,|\,
    2 \mu_1 + \cdots + 2\mu_{l-1} + \mu_l \le k, 
\mu_i \in \Z_{\ge 0} \} .
\ees
The modular transformation matrix $\tilde{S}_{\lambda \mu}$
for $\lambda \in P_+^k(A_{2l-1}^{(2)}), 
\mu \in P_+^k(D_{l+1}^{(2)}),$ takes the form \cite{Kac}
\beq
\label{A:S}
  \tilde{S}_{\lambda \mu} = 
  \frac{i^{l^2}}{\sqrt{2}} (k + 2l)^{-\frac{l}{2}}
  \sum_{w \in W(C_l)} \epsilon(w)
  e^{-\frac{2\pi i}{k+2l}
  (w(\bar{\lambda} + \bar{\rho}), \tilde{\mu} + \tilde{\rho})} .
\eeq
Here the sum is taken over the Weyl group of $C_l$, which is the horizontal
subalgebra of $A_{2l-1}^{(2)}$. 
We denote by $\bar{\lambda}$ the finite part of $\lambda$, which is
expressed by the fundamental weights $\Lambda_i$ of $C_l$ as follows
\beq
  \bar{\lambda} = \lambda_1 \Lambda_1 + \cdots + \lambda_l \Lambda_l .
\eeq
$\tilde{\mu}$ is a weight of $C_l$ determined from $\mu$ via
\beq
\label{def_mu}
  \tilde{\mu} = 2 \mu_1 \Lambda_1 + \cdots + 2 \mu_{l-1} \Lambda_{l-1}
  + \mu_l \Lambda_l . 
\eeq
The Weyl vectors, $\bar{\rho}$ and $\tilde{\rho}$, are defined as
\bes
  \bar{\rho} &= \Lambda_1 + \cdots + \Lambda_l , \\
  \tilde{\rho} &= 2 \Lambda_1 + \cdots + 2 \Lambda_{l-1} + \Lambda_l .
\ees

The outer automorphism group of $A_{2l-1}^{(2)}$ is $\Z_2$.
The generator $A \in \auto(A_{2l-1}^{(2)})$ acts on
$\lambda \in P_+^k(A_{2l-1}^{(2)})$ as
\bes
  A : (\lambda_1,\lambda_2,\cdots,\lambda_{l}) \mapsto &
  (\lambda_0,\lambda_2,\cdots,\lambda_{l}) , \\
  & \lambda_0 = k - (\lambda_1 + 2\lambda_2 + \cdots + 2\lambda_l) .
\ees
One can show this can be written in the form
\beq
  \overline{A\lambda} = k \Lambda_1 + w_A (\bar{\lambda}) ,
\eeq
where $w_A \in W$ is an element of the Weyl group for which
$\epsilon(w_A) = -1$.
Applying this formula to $\lambda + \rho$, we obtain
\beq
  \overline{A\lambda + \rho} =
  (k + 2l) \Lambda_1 + w_A (\bar{\lambda} + \bar{\rho}) ,
\eeq
where we used $A \rho = \rho$. 
Substituting this to \eqref{A:S} yields 
\beq
  \tilde{S}_{A \lambda, \mu} = 
  \frac{i^{l^2}}{\sqrt{2}} (k + 2l)^{-\frac{l}{2}}
  \sum_{w \in W(C_l)} \epsilon(w)
  e^{-\frac{2\pi i}{k+2l}
  (w w_A (\bar{\lambda} + \bar{\rho}), \tilde{\mu} + \tilde{\rho})} 
  e^{- 2\pi i (w \Lambda_1, \tilde{\mu} + \tilde{\rho})} .
\eeq
From the definition of $\tilde{\mu}$, it can be shown that
\beq
  (w \Lambda_1, \tilde{\mu}) = \frac{\mu_l}{2}  \mod \Z \quad
  \text{for any } w \in W,
\eeq
and we obtain the result
\bes
  \tilde{S}_{A \lambda, \mu} &= 
  \frac{i^{l^2}}{\sqrt{2}} (k + 2l)^{-\frac{l}{2}}
  \sum_{w \in W(C_l)} \epsilon(w)
  e^{-\frac{2\pi i}{k+2l}
  (w (\bar{\lambda} + \bar{\rho}), \tilde{\mu} + \tilde{\rho})} \cdot
  \epsilon(w_A) 
  e^{- 2\pi i \frac{\mu_l + 1}{2}} \\
  &= \tilde{S}_{\lambda \mu} (-1)^{\mu_l} .
\ees

The outer automorphism of $D_{l+1}^{(2)}$ is $\Z_2$ and generated by
$\tilde{A}$
\bes
  \tilde{A} : (\mu_1,\cdots,
  \mu_{l-1},\mu_{l}) \mapsto &
  (\mu_{l-1},\mu_{l-2},\cdots,
  \mu_{1},\mu_{0}) , \\
  & \mu_0 =
  k - (
  2\mu_1 + \cdots +
  2\mu_{l-1}+\mu_{l}) .
\ees
The action on $\tilde{S}$ can be calculated in the same way as above.
The result reads
\beq
  \tilde{S}_{\lambda, \tilde{A}\mu} = 
  (-1)^{\lambda_1 + 2 \lambda_2 + \cdots + l \lambda_l}
  \tilde{S}_{\lambda \mu} .
\eeq

%%%%%%%%%%%%%%%%%%%%%%%%%%%%%%%%%%%%%%%%%%%%%%%%%%%%%%%%
%%%%%%%%%%%%%%%%%%%%%%%%%%%%%%%%%%%%%%%%%%%%%%%%%%%%%%%%

%%%%%%%%%%%%%%%%%%%%%%%%%%%%%%%%%%%%%%%%%%%%%%%%%%%%%%%%
%%   References
%%%%%%%%%%%%%%%%%%%%%%%%%%%%%%%%%%%%%%%%%%%%%%%%%%%%%%%%

\end{document}